\documentclass[default]{sn-jnl}
\jyear{2025}%

\theoremstyle{thmstyleone}%

\theoremstyle{thmstyletwo}%

\theoremstyle{thmstylethree}%

\begin{document}

\title{On-chip integrated light sources with MoS\textsubscript{2}/WSe\textsubscript{2} moir{\'e} superlattices at telecom wavelengths}


\author[1]{\fnm{Xingyu} \sur{Huang}}\email{xinhu@dtu.dk}
\author*[1,2]{\fnm{Hanlin} \sur{Fang}}\email{hanfan@dtu.dk}
\author[2,3]{\fnm{Shima} \sur{Kadkhodazadeh}}\email{shka@dtu.dk}
\author[2,3]{\fnm{Monia} \sur{Runge Nielsen}}\email{moniarn@dtu.dk}
\author[1,2]{\fnm{Qiaoling} \sur{Lin}}\email{linqling92@outlook.com}
\author[4]{\fnm{Zhipei} \sur{Sun}}\email{zhipei.sun@aalto.fi}
\author[1,2]{\fnm{Martijn} \sur{Wubs}}\email{mwubs@dtu.dk}
\author*[1,2]{\fnm{Sanshui} \sur{Xiao}}\email{saxi@dtu.dk}

\affil[1]{\orgdiv{Department of Electrical and Photonics Engineering}, \orgname{Technical University of Denmark}, \orgaddress{\postcode{DK-2800}, \state{Kongens Lyngby}, \country{Denmark}}}
\affil[2]{\orgdiv{NanoPhoton - Center for Nanophotonics}, \orgname{Technical University of Denmark}, \orgaddress{\postcode{DK-2800}, \state{Kongens Lyngby}, \country{Denmark}}}
\affil[3]{\orgdiv{National Centre for Nano Fabrication and Characterization Nanocharacterization}, \orgname{Technical University of Denmark}, \orgaddress{\postcode{DK-2800}, \state{Kongens Lyngby}, \country{Denmark}}}
\affil[4]{\orgdiv{Department of Electronics and Nanoengineering and QTF Centre of Excellence}, \orgname{Aalto University}, \orgaddress{\city{Espoo}, \postcode{02150}, \country{Finland}}}

\abstract{On-chip integrated light sources are essential for photonic integrated circuits, requiring waveguides to interface various components, from light sources to detectors. Two-dimensional (2D) transition metal dichalcogenide (TMD) heterostructures offer exceptional tunability and direct bandgaps, opening new avenues for on-chip light sources. However, a waveguide-integrated light source based on 2D materials operating in the telecom windows has yet to be realized. In this work, we demonstrate that the creation of a moir{\'e} superlattice enables light emission in the optical fiber communication (OFC) O-band (1260-1360\,nm) with brightness surpassing that of intralayer excitons in monolayer MoTe\textsubscript{2}. Furthermore, we realize waveguide-integrated light sources emitting in the O-band by integrating these superlattices with asymmetric nanobeam cavities. This cavity design not only significantly enhances light emission but also improves the spectral purity of the single cavity mode. Moreover, the device output remains remarkably stable across varying pump power in an ambient environment, demonstrating excellent operational stability. The device performance remains unchanged over a nine-month measurement period, highlighting its long-term stability. This work presents a new architecture for on-chip light sources, advancing practical photonic applications.}




\maketitle

\unnumbered 
\section{Introduction}\label{}
On-chip integrated light sources are one of the core components for photonic integrated circuits, requiring easy integration of efficient light-emitting materials into existing photonic platforms. Monolayer transition metal dichalcogenides (TMDs) feature a direct bandgap, enabling efficient light emission\,\cite{Splendiani2010}. Additionally, their lack of dangling bonds facilitates seamless integration with various material platforms, e.g., silicon nitride and silicon\,\cite{xia2014two}. TMDs also exhibit unique valley polarization properties, which offer additional degrees of freedom for light manipulation\,\cite{schaibley2016valleytronics}. Furthermore, different TMD monolayers have been used to realize exciton-polariton lasing\,\cite{zhao2021ultralow}, photon lasing\,\cite{shang2017room, li2017room, zhao2021strong}, and light-emitting diodes\,\cite{bie2017mote2, zhu2018high, PalaciosBerraquero2016, gu2019room}. These advantages make monolayer TMDs a promising candidate for developing on-chip integrated light sources, including both classical and quantum light sources. However, the natural bandgap of TMD monolayers limits their applications in technologically important optical fiber communication (OFC) bands. 

In vertically stacked TMD heterostructures, the type-II band alignment leads to the formation of spatially separated yet bound electron-hole pairs, known as interlayer excitons (IXs)\,\cite{fang_strong_2014, rivera_observation_2015}. The diverse combinations of TMD monolayers enable IX emissions with tunable wavelengths on demand\,\cite{jiang2021interlayer}. Moreover, moir{\'e} superlattices created in twisted TMD heterostructures can effectively engineer the band structure, tailoring IX emissions\,\cite{huang2022excitons, lin_moire-engineered_2024} and enabling the exploration of novel exciton physics\,\cite{yu2017moire, regan2022emerging}. These features highlight the potential of IXs for photonic applications\,\cite{du2023moire}. Specifically, IX-based lasers coupled to free space have been demonstrated using photonic crystal cavities\,\cite{liu2019room, Paik2019}. Further efforts have been directed toward developing IX-based on-chip light sources operating around 920\,nm\,\cite{khelifa_coupling_2020}. However, a waveguide-integrated light source at OFC bands using IXs remains unexplored. In this context, while efficient IX emission at room temperature is highly desirable, the photoluminescence (PL) of IXs is typically weaker than that of intralayer excitons\,\cite{tran2019evidence, Sun2022}, underscoring the need for enhanced light emission efficiency. 

In this work, we improve the light emission efficiency of IXs by creating a moir{\'e} superlattice in a MoS\textsubscript{2}/WSe\textsubscript{2} heterobilayer with a near-zero twist angle. The remarkable efficiency is validated by comparing the light emission of IXs to that of intralayer excitons in monolayer MoTe\textsubscript{2}, a well-established direct bandgap semiconductor\,\cite{ruppert_optical_2014}. Through PL and time-resolved spectroscopy, we observe that at room temperature, the PL intensity of IXs can exceed that of intralayer excitons in monolayer MoTe\textsubscript{2} while retaining an exciton lifetime that is an order of magnitude longer. This enhancement is attributed to the formation of moir{\'e} potential, which effectively suppresses non-radiative recombination. Further, we integrate the superlattices with asymmetric silicon nanobeam cavities. The resulting devices exhibit narrow single-mode emission in the OFC O-band (1260-1360\,nm) at low pump powers, with a high side-mode suppression ratio (SMSR) and efficient waveguide coupling. Additionally, the device performance dose not degrade over the nine month measurement period. These findings open new avenues for integrated photonic light sources, particularly in silicon photonics.

\section{Results}\label{}
\subsection{Bright IX emission}\label{}

\begin{figure}[h]
\centering
\includegraphics[scale=0.7]{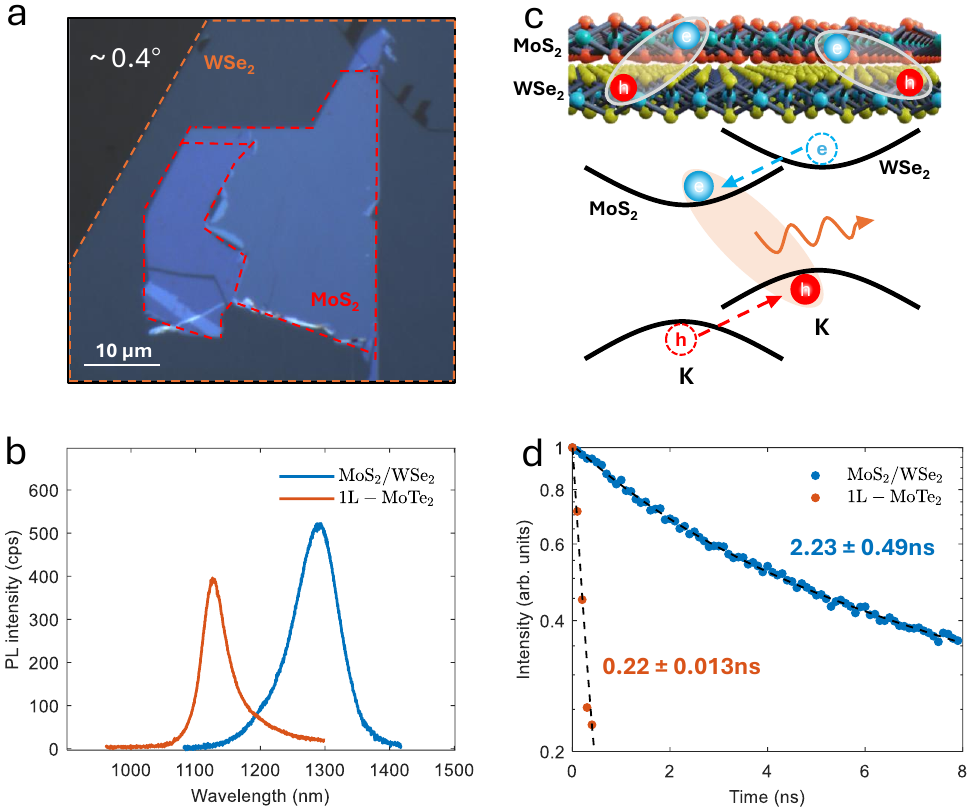}
\caption{Light emission of a MoS$_2$/WSe$_2$ heterobilayer. (a) Optical microscope image of a MoS$_2$/WSe$_2$ heterobilayer with a twist angle of $\sim$0.4$^\circ$, showing two distinct heterobilayer regions identified by different optical contrasts. (b) PL spectra of IXs in the heterobilayer and intralayer excitons in a 1L-MoTe\textsubscript{2}. (c) Schematic illustration of IXs in a heterobilayer with type-II band alignment (top) and the band diagram of the heterobilayer at the K valley (bottom). The dashed arrows represent the movement of charge carriers. (d) PL decay curves of IXs in the heterobilayer and intralayer excitons in a 1L-MoTe\textsubscript{2}. 
}
\label{fig1}
\end{figure}

The MoS\textsubscript{2} and WSe\textsubscript{2} monolayers are prepared on polydimethylsiloxane (PDMS) substrates and then stacked (details in Methods) to obtain a high-quality interface. Fig.~\ref{fig1}a presents an optical microscope image of the fabricated MoS\textsubscript{2}/WSe\textsubscript{2} heterobilayer without the widely observed bubbles\,\cite{baek2020highly, rodriguez2021strong, lin_moire-engineered_2024}, which weaken interlayer coupling and reduce IX emission efficiency. The stacking process results in the heterobilayer of Fig.~\ref{fig1}a having two distinct regions. The observed strong PL emission (see Fig.~\ref{fig1}b) originates from the right region, while it is not observed for the left region. This stark difference is due to the different interlayer coupling for the two regions, which is supported by the PL quenching of intralayer excitons in the constituent WSe$_2$ monolayer (see Fig.~S1). We note that this emission originates from IXs formed through rapid charge transfer between monolayers (see Fig.~\ref{fig1}c)\,\cite{chen2016ultrafast}, strongly depending on interlayer coupling. The energy of electrons and holes in the K valleys regulates the IX emission wavelength. Here, the emission wavelength is extended to the OFC O-band for the heterobilayer with a near-zero twist angle ($\sim$0.4$^\circ$). 

To identify whether radiative recombination of IXs is efficient at room temperature, we compare their PL spectrum with that of intralayer excitons in 1L-MoTe\textsubscript{2} under identical measurement conditions. 1L-MoTe\textsubscript{2} is expected to exhibit brighter PL emission than that of IXs, which consist of electrons and holes in separate monolayers; however, this is found not to be the case for our experiment. To the contrary, Fig.~\ref{fig1}b shows that the PL emission intensity of IXs in the MoS\textsubscript{2}/WSe\textsubscript{2} heterobilayer is stronger than that of intralayer excitons in 1L-MoTe\textsubscript{2}. Notably, some other heterobilayers exhibit a peak PL intensity up to four times higher than that of 1L-MoTe\textsubscript{2} (see Fig.~S2). The similar absorption coefficients of MoS\textsubscript{2}, WSe\textsubscript{2}, and MoTe\textsubscript{2}\cite{ruppert_optical_2014, li_measurement_2014} help us rule out that the photo-excited carrier densities at the excitation wavelength of 640\,nm would differ much. Moreover, it should be noted that 1L-MoTe\textsubscript{2} is freshly exfoliated for PL measurements, avoiding the effect of sample degradation over time reported in previous work\,\cite{fang2019laser}. 

Furthermore, through time-resolved PL measurements, we find that the IXs show a lifetime that is one order of magnitude longer than that of the intralayer exciton in 1L-MoTe\textsubscript{2}. We note that the exciton lifetime of MoTe\textsubscript{2} should be a few picoseconds\,\cite{robert2016excitonic} (see Fig.~\ref{fig1}d). The measured value of 0.22\,ns is limited by the resolution of our instrument. Typically, for emitters with the same quantum yield, a bright emitter would show a faster PL decay. Therefore, the observation of brighter IX emission with a much longer exciton lifetime reflects a higher PL quantum yield of the heterobilayer than that of 1L-MoTe\textsubscript{2}. Decrasing twist angle has been found beneficial for the creation of atomic reconstruction, leading to a periodical strain distribution with enhanced moir{\'e} potential\,\cite{lin_moire-engineered_2024}. The resulting moir{\'e}  potential traps can hinder the transport of IXs\,\cite{choi2020moire, yuan2020twist, li2021interlayer}, thereby reducing the likelihood of their capture by non-radiative centers. Here, the creation of a moir{\'e} superlattice is visualized by transmission electron microscopy (TEM) measurements (see Fig.~S3 and Methods for details), and the existence of the moir{\'e} potential is further confirmed by the observation of multiple IX states\,\cite{tran2019evidence, tan2022signature, lin_moire-engineered_2024} at high pump powers (see Fig.~S4). Therefore, we attribute the bright IX emission and long exciton lifetime to moir{\'e}-suppressed non-radiative recombinations. The high PL efficiency and O-band emission wavelengths of IXs in moir{\'e} superlattices make them promising for silicon-based light-emitting devices.

\unnumbered 
\subsection{Asymmetric nanobeam cavity}\label{}

\begin{figure}[h]
\centering
\includegraphics[scale=0.7]{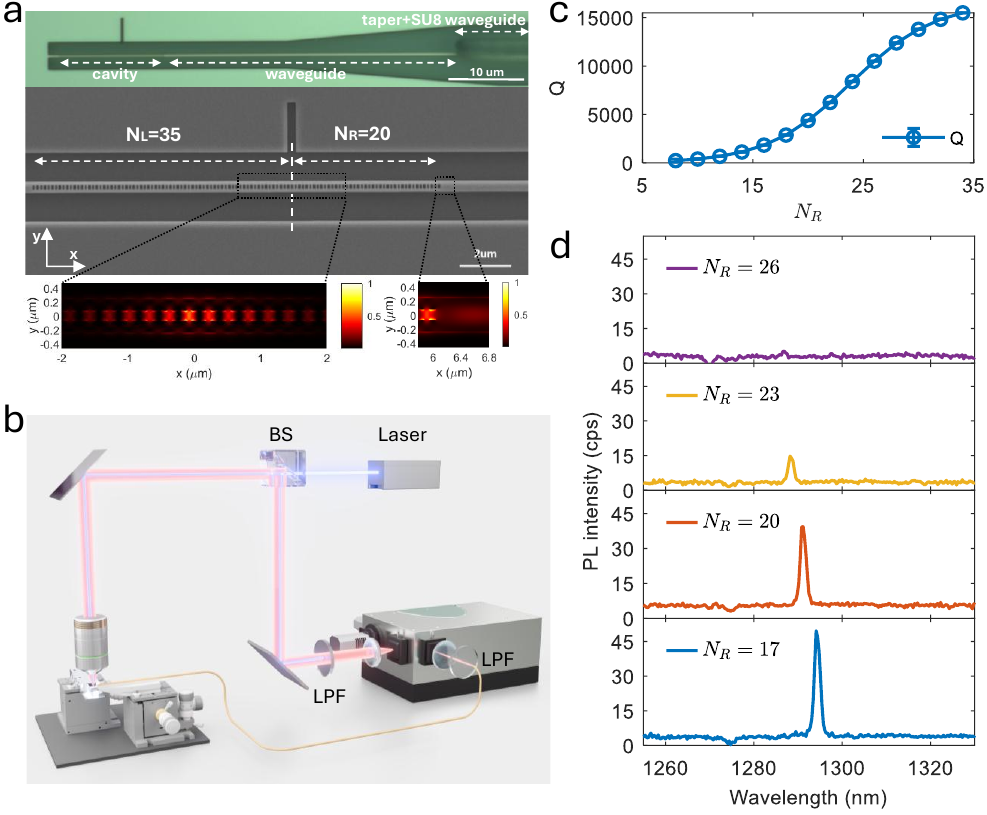}
\caption{Asymmetric silicon nanobeam cavity. (a) Optical microscope image of the cavity (top), SEM image of the cavity (middle), and simulated electric-field profile in the \textit{xy}-plane (bottom). (b) PL measurement setup for both free-space and waveguide-coupling characterization. BS: 50:50 beam splitter, LPF: long-pass filter. (c) Relation between the simulated cavity Q-factor and the number of periods N$_{R}$ of the Bragg mirror on the right. (d) Silicon-PL-excited cavity mode signal at different N$_{R}$ values, acquired via waveguide coupling.
}
\label{fig2}
\end{figure}

To explore integrated light sources using MoS\textsubscript{2}/WSe\textsubscript{2} moir{\'e} superlattices, an asymmetric silicon photonic crystal nanobeam cavity supporting waveguide coupling is designed (see Fig.~\ref{fig2}a for the fabricated device and Fig.~S5 for the design parameters). The number of unit cells on the right side of the cavity (denoted as N$_{R}$, see Fig.~\ref{fig2}a for the SEM image) is smaller than that on the left side, and the cavity is connected to a silicon waveguide. This design allows us to achieve practically one-sided on-chip emission from the asymmetric cavity into the waveguide. Additionally, a silicon taper and an SU-8 waveguide are employed to achieve high broadband butt-coupling efficiency\,\cite{pu_ultra-low-loss_2010, keller_processing_2008}. 

The signal from the SU-8 waveguide is collected by a lensed fiber and connected to a spectrometer (see Fig.~\ref{fig2}b). The bottom inset of Fig.~\ref{fig2}a shows the localized resonant mode around the cavity center and the evanescent coupling into the silicon waveguide. The cross-polarization scattering method\,\cite{deotare2009high} is typically used to detect the optical modes of a photonic cavity. However, it becomes ineffective for our cavities, which are not underetched. Instead, we characterize the cavity mode by measuring silicon PL under a relatively high optical pump power (up to 5\,mW) because of the indirect bandgap of silicon. The existence of the cavity mode with good waveguide coupling is confirmed by the observation of a single emission peak. We note that the precise evaluation of the cavity Q-factor is largely limited by the spectral resolution of the grating spectrometer. A 150 grooves/mm grating is used to detect the cavity mode excited by the weak PL signal of silicon. Therefore, the measured Q-factors are lower limits of the true values of the cavities. With this cavity design, increasing N$_R$ enhances the cavity Q-factor, while the waveguide-coupled PL signal correspondingly decreases. Fig.~\ref{fig2}c illustrates the dependence of the cavity Q-factor on N$_R$, with details provided in the Methods section. To obtain a feasible coupling efficiency while maintaining a relatively high Q-factor, we strategically reduce N$_{R}$ and employ a large period number on the left side (N$_{L}$=35). $N_R$ = 20 is chosen because it contributes to a relatively high Q-factor (see Fig.~\ref{fig2}c) and an efficient waveguide coupling, evidenced by the remarkable cavity-enhanced PL of silicon (see Fig.~\ref{fig2}d).

\unnumbered 
\subsection{Waveguide-integrated telecom light sources}\label{}

\begin{figure}[h]
\centering
\includegraphics[scale=0.65]{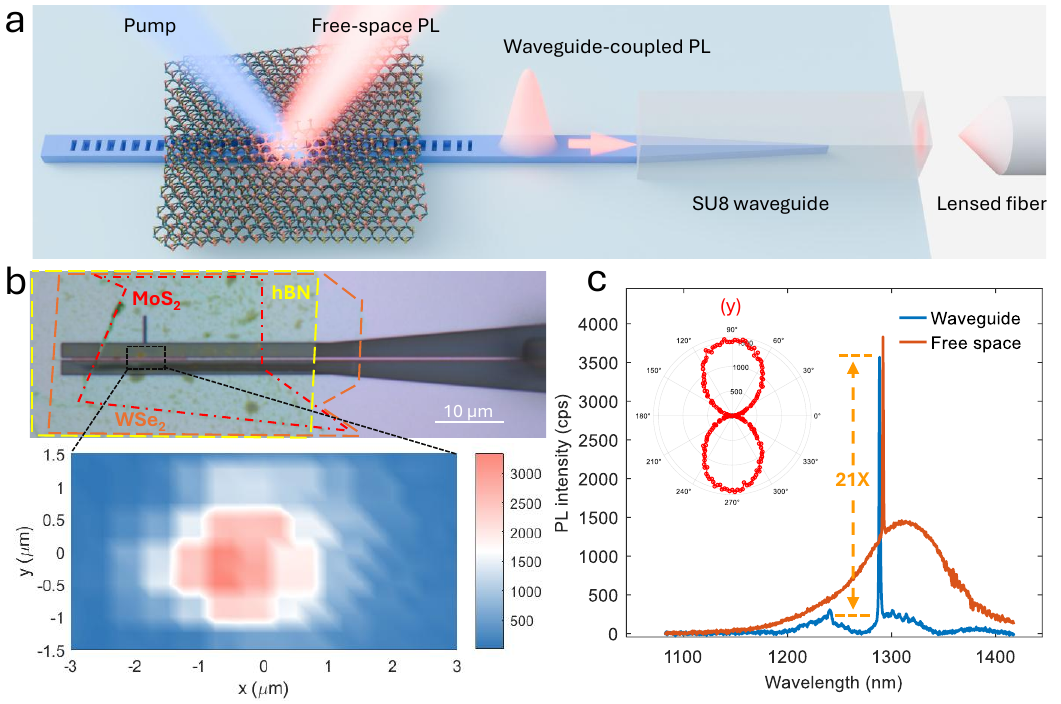}
\caption{The hBN-capped MoS\textsubscript{2}/WSe\textsubscript{2} heterobilayer on an asymmetric cavity. (a) Sketch of the device and collection paths for the PL signal. The hBN layer on top of the heterobilayer is omitted for clarity in the structural illustration. (b) Optical micrograph image of the device. The bottom plot shows the free-space PL mapping of the cavity central region marked by the black dashed rectangle. (c) PL spectrum of the device for free-space and waveguide coupling at 128\,$\mu$W pump power. The single narrow emission peaks confirm the single-mode operation. The inset exhibits the polarization of the cavity mode with a dominant component (y-direction) perpendicular to the nanobeam axis (x-direction).
}

\label{fig3}
\end{figure}

MoS\textsubscript{2}/WSe\textsubscript{2} heterobilayers are then transferred to the top surface of the asymmetric cavity by the van der Waals pickup method (details in the Methods). To improve the mode overlap between IXs and the cavity, only the top hBN layer is adopted rather than the typical sandwich encapsulation structure\,\cite{tran2019evidence, baek2020highly, tan2022signature}. Fig.~\ref{fig3}a and Fig.~\ref{fig3}b shows the schematic diagram and the optical micrograph of the device, respectively. We emphasize that the coverage of the heterostructure does not affect the formation of the cavity mode, which is confirmed by the PL mapping measurements. The bottom plot of Fig.~\ref{fig3}b presents that the cavity-enhanced emission is spatially confined within an area of approximately 1.5\,$\mu$m\,×\,1.5\,$\mu$m at the center of the cavity. 

Fig.~\ref{fig3}c shows the PL spectra obtained via free-space and waveguide coupling. Compared to the free-space PL signal, the spectrum acquired through waveguide coupling exhibits a dramatically lower background (SMSR$\sim$21) while maintaining nearly the same peak intensity at the cavity mode. This contrast indicates that the background in the free-space collection originates from heterobilayer regions that are not efficiently coupled to the cavity. Photons emitted by IXs at wavelengths resonant with the cavity are (i) spatially confined by the cavity mode and (ii) efficiently redirected into the waveguide via evanescent coupling. As a result, our asymmetric cavity design not only enhances the IX PL emission but also improves the spectral purity of the cavity-enhanced PL intensity in the OFC O-band. Similar characteristics are observed in other devices that we fabricated (see Fig.~S6). Furthermore, the device demonstrates strong linear polarization (see the inset of Fig.~\ref{fig3}c and Fig.~S7), suggesting efficient coupling of IX emissions into the cavity mode and highlighting its potential for polarization-sensitive photonic applications, including quantum optics and optical communication systems.

\begin{figure}[h]
\centering
\includegraphics[scale=0.7]{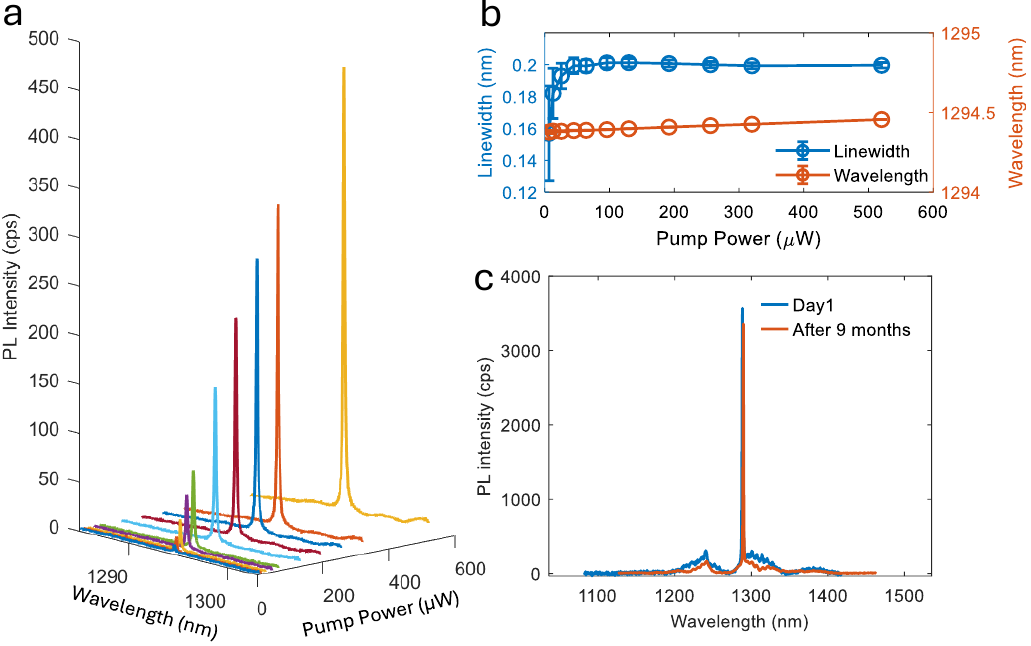}
\caption{Output properties of the device. (a) PL spectrum recorded at different pump powers. (b) Linewidth and center wavelength extracted from (a) as a function of the pump power. (c) Waveguide-coupled signal measured over a 9-month interval, demonstrating excellent stability in the air.
}
\label{fig4}
\end{figure}

To systematically investigate the output properties of the device, we vary the pump power and record the PL spectra using a 1200 grooves/mm grating (see Fig.~\ref{fig4}a). This grating provides a high spectral resolution of $\sim$0.1\,nm for characterizing the emission linewidth of the cavity mode. As the pump power increases, the peak intensity of the cavity mode grows significantly faster than the PL background (see also Fig.~S8), resulting in a high SMSR crucial for optical communication. Notably, the cavity mode centered at 1294\,nm consistently dominates the PL spectra across all pump powers. By fitting the emission peak with a Lorentzian function (see Fig.~S9), we extract the linewidth and wavelength (see Fig.~\ref{fig4}b). A linewidth of $\sim$0.2\,nm is achieved at a low pump power of 6.4\,$\mu$W, which is narrower than what was previously reported for TMD-based lasers\,\cite{zhao2021strong}. Furthermore, both linewidth and wavelength remain nearly constant with increasing pump power, indicating that the device performs well as a heat sink. Importantly, the operating lifetime of the device is another figure of merit for practical device applications. To validate the long-term robustness of the device, the waveguide-coupled PL is monitored over nine months under ambient conditions. As quantified in Fig.~\ref{fig4}c, the variation of the PL intensities is negligible, confirming that the device is robust against air oxidation. This operational resilience positions the superlattice-cavity devices as promising candidates for reliable on-chip light sources.

\section{Discussion}\label{}
In summary, we have demonstrated the advantages of using moir{\'e} superlattices for on-chip light sources, featuring enhanced light emission at telecom wavelengths. The excellent strain tunability of 2D-TMDs\,\cite{cho2021highly} could potentially extend the IX emissions to the OFC C-band (1530–1565\,nm), paving the way for IX-based long-distance optical communication. Furthermore, the cavity design with a small footprint is beneficial for developing ultra-compact light source arrays.

Last but not least, the output intensity and linewidth of the devices show no threshold behavior as a function of pump power. These features suggest that lasing action is not achieved or the threshold is below the detection limit of our instrument. It should be noted that the cavity Q-factor ($\sim$6472) is larger than previously reported for 2D material-based lasers at room temperature\,\cite{li2017room, liu2019room}, suggesting that improving the quantum yield of IXs and understanding the gain mechanism in moir{\'e} superlattices are crucial.

\section{Materials and methods}\label{}
\subsection{Sample fabrication}\label{}
The monolayers of MoTe\textsubscript{2}, MoS\textsubscript{2} and WSe\textsubscript{2} we studied here were exfoliated from flux bulk material (2D Semiconductors) on the PDMS stamps using the standard scotch tape method \cite{Novoselov2005}. The MoS\textsubscript{2}/WSe\textsubscript{2} heterostructure was stacked using an all-dry transfer technique \cite{CastellanosGomez2014}. Care was taken to rotationally
align the MoS\textsubscript{2} and WSe\textsubscript{2} monolayers based on their clear and straight edge under the optical microscope. The MoS\textsubscript{2}/WSe\textsubscript{2} heterostructure was encapsulated on the top with hBN using the van der Waals pick-up method and subsequently transferred to the silicon cavity. This top-side hBN encapsulation was specifically chosen to prevent degradation and contamination while maximizing the coupling efficiency between the MoS$_2$/WSe$_2$ heterostructure and the silicon cavity.

The silicon nanobeam cavity and the inverse taper were fabricated on a 220\,nm silicon-on-insulator (SOI) wafer from Soitec. A 180\,nm layer of CSAR electron beam resist was spin-coated (4000\,rpm, 60\,seconds) on the silicon substrate. The desired pattern was defined in the resist using electron-beam lithography (JEOL 9500, 100\,kV) and subsequently transferred into the 220\,nm silicon layer via reactive ion etching. The residual resist was removed using 1165 solvent, followed by gentle oxygen plasma cleaning. To fabricate the SU-8 polymer waveguides (5\,$\mu$m × 5\,$\mu$m cross-section), a 5\,$\mu$m thick UV photoresist (SU8-2005) was spin-coated onto the existing silicon structures. The pattern was exposed using UV lithography (Maskless Aligner 150, Heidelberg Instruments), developed, and then soft-baked at 90\,℃ for 15\,hours to prevent SU-8 waveguide breakage\,\cite{keller_processing_2008}.

\subsection{Optical spectroscopy}\label{}
All measurements were conducted at room temperature. PL was excited using a 640\,nm continuous-wave (CW) laser (PicoQuant LDH-IB-640B) focused through a 50$\times$ objective lens (NA = 0.65, Olympus LCPLN50XIR), yielding a spot size of $\sim$1.5\,$\mu$m. The emitted signal was filtered by long-pass spectral filters and dispersed using a Czerny-Turner monochromator (Andor SR500i) equipped with a cooled InGaAs array detector (Andor DU491A-1.7). For waveguide-coupled PL characterization, a lensed fiber (3.5\,$\mu$m spot diameter, OZ Optics) was aligned to the SU8 waveguide under the same 50X objective. A 150\,grooves/mm grating was applied for wide-range spectra measurements, and the finer gratings (1200\,grooves/mm) were mainly applied for cavity mode property analysis.  

Time-resolved PL utilized a time-correlated single-photon counting (TCSPC) system. Samples were excited by a 640\,nm pulsed laser (PicoQuant LDH-IB-640B, \(<\)90\,ps pulse width, 10\,MHz repetition rate). The emission signal was fiber-coupled to a home-built filtering system (0.2\,nm bandwidth), whose operational wavelength range was adjusted by rotating the grating. The filtered signal was subsequently directed to a single-photon detector (IDQ ID230).


\subsection{TEM measurements}\label{}
High-angle annular dark-field scanning transmission electron microscopy (HAADF STEM) images were recorded using a Thermo Fisher Scientific Spectra instrument equipped with a field emission electron gun and aberration correction. The TEM instrument was operated at 120\,kV, in order to avoid severe knock-on damage to the sample. The electron probe convergence angle and diameter were approximately 24\,mrad and 170\,pm, respectively, and the HAADF detector collection angle was set to $\sim$45\,mrad. The acquired image (see Fig.~S3) was denoised in the postprocessing using an open-source deep convolutional neural network (CNN) approach by using tk\_r\_em, a python toolkit based on Tensor Flow, which is proposed in Ref.\,\cite{Lobato2024}.

\subsection{Numerical simulation}\label{}
The photonic cavity design adopts a topological nanobeam architecture, first introduced in the pioneering work\,\cite{ota_topological_2018}. Building on our previous design\,\cite{fang2019laser, lin_moire-engineered_2024}, we retained the essential SiO\textsubscript{2} substrate layer, which ensures both mechanical stability for van der Waals material integration and enhanced heat dissipation capability. Using finite-difference time-domain (FDTD) simulations (Lumerical, Ansys Inc.), we systematically optimized the design by sweeping N$_R$ and recording the Q-factor to achieve a good trade-off between Q-factor and waveguide coupling efficiency.

\bmhead{Acknowledgments}
This work is partly supported by the Danish National Research Foundation through NanoPhoton – Center for Nanophotonics (Grant Number DNRF147). X.H., H.F., and S.X.~acknowledge the support from the Independent Research Fund Denmark (2032-00351B) and Villum Fonden (VIL70202). M.R.N.~acknowledges the support from Villum Fonden (VIL58716).

\bmhead{Author contributions}
H.F.~and S.X.~conceived the project. H.F.~and X.H.~designed the device. X.H.~and Q.L.~fabricated the devices. X.H.~performed optical characterizations. H.F.~and X.H.~carried out simulations of the asymmetric cavities. S.K.~performed the TEM measurements. S.K.~, M.R.N., and H.F.~analyzed the TEM data. X.H., H.F., Z.S., M.W., and S.X.~analyzed the PL measurement results. All authors contributed to the discussion and writing of the manuscript.

\bmhead{Data availability}
All data necessary to evaluate the conclusions are provided in the paper. Additional data related to this study are available from the corresponding authors upon reasonable request.

\bmhead{Conflict of interest}
The authors declare no conflict of interest.

\clearpage


\end{document}